# Quantum Storage of Orbital Angular Momentum Entanglement in an Atomic Ensemble

Dong-Sheng Ding[1,2], Wei Zhang[1,2], Zhi-Yuan Zhou[1,2], Shuai Shi[1,2], Guo-Yong Xiang[1,2], Xi-Shi Wang[3], Yun-Kun Jiang[4], Bao-Sen Shi[1,2,*] and Guang-Can Guo[1,2]

[1]*Key Laboratory of Quantum Information, University of Science and Technology of China, Hefei, Anhui 230026, China and*

[2]*Synergetic Innovation Center of Quantum Information & Quantum Physics, University of Science and Technology of China, Hefei, Anhui 230026, China*

[3]*State Key Lab. of Fire Science, University of Science & Technology of China, Hefei, Anhui 230026, China*

[4]*College of Physics and Information Engineering, Fuzhou University, Fuzhou, 350002, P. R. China*

*Corresponding author: \*drshi@ustc.edu.cn*

Constructing a quantum memory for a photonic entanglement is vital for realizing quantum communication and network [1-4]. Besides enabling the realization of high channel capacity communication [5], entangled photons of high-dimensional space are of great interest because of many extended applications in quantum information and fundamental physics fields [6-9]. Photons entangled in a two-dimensional space had been stored in different system [10-13], but there have been no any report on the storage of a photon pair entangled in a high-dimensional space. Here, we report the first experimental realization of storing an entangled orbital angular momentum (OAM) state through a far off-resonant two-photon transition (FORTPT) in a cold atomic ensemble. We reconstruct the matrix density of an OAM entangled state postselected in a two-dimensional subspace with a fidelity of 90.3%±0.8% and obtain the Clauser, Horne and Shimony and Holt inequality parameter $S$ of 2.41±0.06 after a programmed storage time. All results clearly show the preservation of entanglement during the storage. Besides, we also realize the storage of a true-single-photon via FORTPT for the first time.

The establishment of quantum network in the future needs distribution of quantum entangled photons over channels between different nodes [14, 15]. To overcome the exponential scaling of the error rate with the channel length, the concept of quantum repeater is introduced [16], which combines entanglement swapping and quantum memory to efficiently extend the achievable distance of quantum communication. During the last years, important progresses have been made towards the realization of an efficient and coherent quantum memory based on gas and solid atomic ensemble [17-21], photons encoded in a two-dimensional space spanned for example by orthogonal polarizations or different paths had been stored [10-13]. Moreover, many groups and researchers are active in storing light encoded using a high-dimensional space in different physical systems [22-32]. In quantum information and quantum optics fields, a photon encoded in a high-dimensional space [33-36] could carry $\log_2 d$ bits information, where $d$ is the number of orthogonal basis vectors of the Hilbert space. In such a way, the transmission rate of quantum communications is increased greatly [37], and the capacity of channel could be also significantly improved [5]. Moreover, it affords quantum key distribution a more secure flux of information [38], etc. Because of the inherent infinite dimension of orbital angular momentum (OAM) space[39-41], a light is usually encoded in OAM space to offer the higher-information-density coding. Therefore, the preparation of a high-dimensional OAM entangled state plays a vital role inquantum information and communication fields, and usually was realized by using the spontaneous parametric down-conversion in a crystal [41] or spontaneous Raman scattering (SRS) in an atomic ensemble [42, 43] experimentally.

Building up a quantum network based on OAM involves the coherent interaction [3] between OAM entangled photons and the matter, so storing an entanglement of OAM state is critical for establishing a high-dimensional quantum memory. The present implemented memory protocols include electromagnetically induced transparency, far off-resonant two-photon transition (FORTPT), controlled reversible inhomogeneous broadening, and atomic frequency combs. Of which, FORTPT protocol has some interesting features, such as the ability to store a broadband pulse towards high-speed quantum memories, the insensitivity to inhomogeneous broadening, the entangling the memory with the transmitted optical mode, etc. [44]. Walmsley's group experimentally realized the storage of a light near single-photon level via FORTPT recently [44, 45], however, still it is strongly attenurated laser, not a true single photon. Besides, there have been no any report on storing a light carrying an image or OAM via FORTPT. Furthermore, since Zeilinger's group observed the entangled properties of OAM [41] in 2001 year, there have been no any experimental progresses for storing entangled OAM states via any protocol in any physical memory system. Experimentally realizing the storage of the OAM entanglement is a big challenge.

Here, we report the first experimental realization of a quantum memory for an OAM entanglement. In the experiment, we establish the OAM entanglement between the anti-Stokes photon and the collective spin excited state of the atomic ensemble in one cold atomic ensemble by SRS firstly, this entanglement is a high-dimensional OAM entanglment [42, 43]. Then we send this anti-Stokes photon to and store it via FORTPT in another cold atomic ensemble acting as the quantum memory. By this way, a high-dimensional entanglement is established between two atomic ensembles. We could demonstrate this OAM entanglement by mapping the spin excited states in two ensembles to two photons and checking their entanglement. In our experiment, we prove the entanglement existed in a two-dimensional subspace for experimental simplicity. We reconstruct the density matrix of the OAM entangled photon pair with a fidelity of 90.3%±0.8% and obtain the Clauser, Horne and Shimony and Holt (CHSH) inequality parameter of $S$=2.41±0.06 after a programmed storage time. All results clearly show the preservation of the entanglement in our gas memory system. The work here includes three significant progresses: 1. the first experimental realization of storing an entangled OAM state; 2. the first experimenal storage of a true single photon via FORTPT; 3. the first experimental evidence of a true single-photon-carrying OAM stored via FORTPT.

**Experimental Results**

The layout of our experiment was depicted by Fig. 1. In the experiment, pump 1 was applied firstly for generating an anti-Stokes photon at 795 nm called signal 1 by a non-colinear SRS process in an optically thick Rubidium (Rb) ensemble, which was trapped in a two-dimensional magneto-optical trap (MOT1) [46]. By using a series of mirrors and lenses (a 4-f imaging system was consisted), the generated signal 1 photon was delivered into the second MOT 2 for subsequent storage. A coupling pulse laser with orthogonal polarization to signal 1 was used to store the signal 1 via FORTPT (see Methods). After the signal 1 photon was retrieved from MOT 2, the pump 2 laser was switched on for converting the collective spin excited state of the atomic ensemble in MOT 1 to a Stokes photon at 780 nm called signal 2.

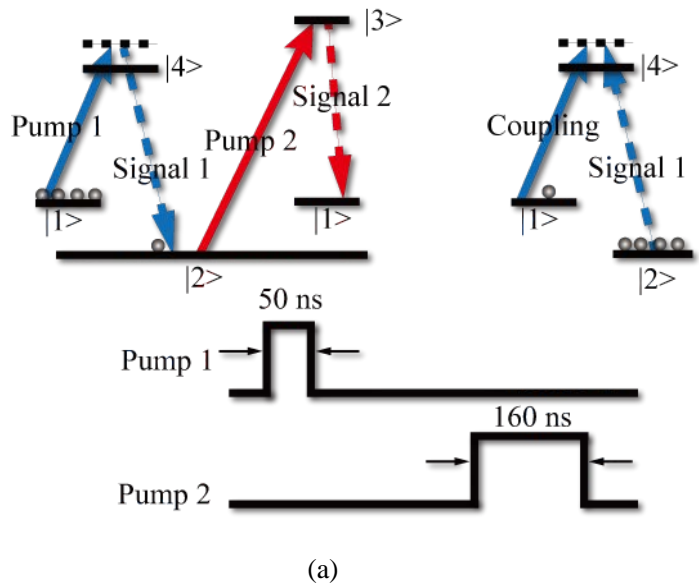

(a)

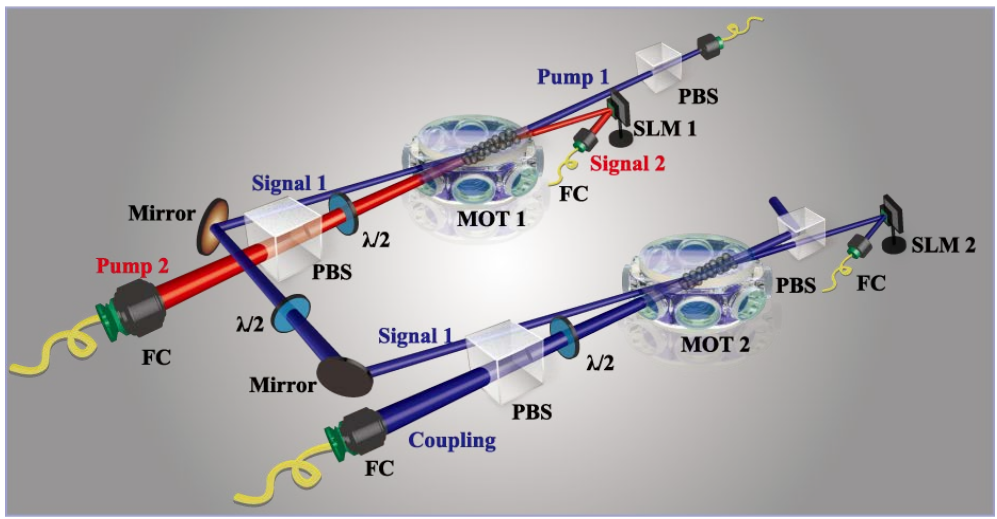

(b)

**Figure.1|Simplified energy level diagram and experimental setup.** (a). Simplified energy level diagram of the SRS. The states |1> and |2> correspond to two metastable levels $5S_{1/2}(F=3)$ and $5S_{1/2}(F=2)$ of $^{85}$Rb atom respectively, |3> and |4> are the excited levels of $5P_{3/2}(F'=3)$ and

$5P_{1/2}$(F′=3) respectively. The pump 1 laser is from an external-cavity diode laser (DL100, Toptica) with the wavelength of 795 nm, and is blue-detuned to the atomic transition of $5S_{1/2}$(F=3)->$5P_{1/2}$(F′=3) with a value of 70 MHz. The pump 2 laser is from another external-cavity diode laser (DL100, Toptica) with the wavelength of 780 nm which couples the atomic transition of $5S_{1/2}$(F=2) ->$5P_{3/2}$(F′=3). The pump 1 and pump 2 are modulated into pulse modes with a width of 50 ns and 160 ns respectively, and a rising edgeof 30 ns. The delayed time between the pump1 pulse and the pump 2 pulse is programed to be 260 ns for the process of storage. The powers of pump 1 and pump 2 are 0.5 mW and 4 mW respectively. The coupling laser is from the same laser with pump 1 and is also blue-detuned to atomic transition of $5S_{1/2}$(F=3)->$5P_{1/2}$(F′=3) with a value of 70 MHz, its power is about 12 mW. (b). Simplified diagram depicting the storage of entanglement of OAM state. The waist of signal 1 at MOT 2 was 63 μm. MOT: magneto-optical trap; FC: fibre coupler; SLM: spatial light modulator; PBS: polarisation beam splitter; λ/2: half-wave plate.

Firstly, we established the nonclassical correlation in time domain between atomic ensembles in MOT 1 and MOT 2 by storing the signal 1 photon in MOT 2. In this case, the delayed time between the pump 1 and pump 2 pulses was programed to be 260 ns, both SLMs were set to act as mirrors. In our experiment, before the pump 2 was applied, we added a coupling laser to store the generated signal 1 photon via FORTPT in MOT 2 for a while. By this way, we built up the non-classical correlation between two MOTs. We could demonstrate this nonclassical correlation in time domain by mapping the spin excited states in two ensembles to two photons and checking their correlation. After had retrieved signal 1 photon in MOT 2, we used the pump 2 laser to map the collective spin excited state of the atomic ensemble in MOT 1 to signal 2 photon. We measured the cross-correlation function $g_{s1,s2}(\tau)$ against the storage time, the results were shown in Fig. 2. We first proved the existence of a non-classical correlation between these two photons by demonstrating a strong violation of the Cauchy-Schwarz inequality (see Methods). Furthermore, we also demonstrated the single-photon property of the signal 1 photon by performing the Hanbury-Brown and Twiss (HBT) experiment on the trigger photon. Experimentally, we obtained an $\alpha$ value of 0.074±0.0012 before storage and 0.29±0.02 after 150 ns-long storage, confirming clearly that the single-photon nature is preserved during storage. For an ideally prepared single-photon state, the anti-correlation parameter $\alpha$ goes to zero. Details of the experiment are presented in the Methods.

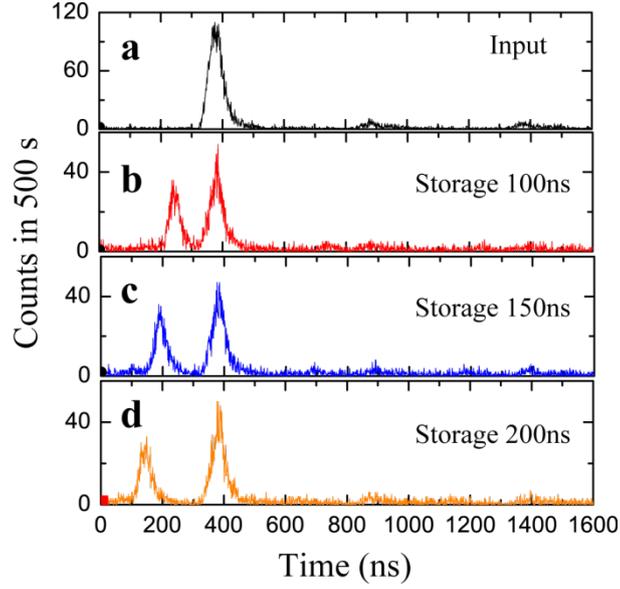

**Figure.2|.The measurement of cross-correlated function g$_{s1,s2}$(τ) in the process of storage.**(a) cross-correlated function g$_{s1,s2}$(τ) between signal 1 and signal 2 photons with a delayed time of 260 ns between pump 1 and pump2.(b), (c) and (d) were the time-correlated function g$_{s1,s2}$(τ) between signal 2 photon and the retrieval signal 1 photon with storage time of 100 ns, 150 ns and 200 ns respectively.The signal 1 acted as trigger photon, and the signal 2 acted as stop signal.The spatial modes of both signal 1 and signal 2 were Gaussian. All data were raw, without noise correction.

Next we moved to the main part of this work: establishing an OAM entanglement in two atomic ensembles by storing signal 1 photon in MOT 2. In the experiment, we built tup the OAM entanglement between the anti-Stokes photon and the collective spin excited state of the atomic ensemble in one cold atomic ensemble in MOT 1 by SRS firstly, this entanglement was specified by the formula of $|\psi\rangle = \sum_{l=-\infty}^{l=\infty} c_l |l\rangle_{s1} \otimes |-l\rangle_{a2}$ [42, 43], is a high-dimensional entangled state. Where subscripts s1 and a2 labeled the signal 1 photon and the atomic ensemble in MOT 1 respectively, $|c_l|^2$ is the excitation probability, $|l\rangle$ is the OAM eigenmode with quanta of *l*. After we stored signal 1 photon in the atomic ensemble in MOT 2, a high-dimensional entangled state between two atomic ensembles was established. We could demonstrate this OAM entanglement by mapping the spin excited states in two ensembles to two photons and checking their entanglement. But in order to simplify the experiment, shorten the measurement time, here we only experimentally demonstrated the storage of the OAM entanglement postselected in a two-dimensional subspace. In order to characterize the obtained state, two-qubit state tomography [47] was performed to reconstruct the density matrix of state. In the experiment, we set the delayed time of the applied pump 2 to be 260 ns. Before pump 2 was applied, the OAM entanglement between signal 1 photon and the ensemble in MOT 1 was established, which could be verified by mapping the OAM of the ensemble in MOT 1 to signal 2 photon, and checking OAM entanglement between signal 1 and signal 2 photons. By projecting signal 1 and signal 2 photons on basis vectors of $|L>$, $|R>$, $(|L>-i|R>)/2^{1/2}$, $(|L>+|R>)/2^{1/2}$ (see Methods) by using two SLMs, we obtained corresponding 16 coincidence rates, then used them to reconstruct the density matrix of the state. Fig. 3(a) and Fig. 3(b) were the corresponding real and imaginary parts of the reconstructed density matrix. By using the formula of $F_1 = Tr(\sqrt{\sqrt{\rho_{input}}\rho_{ideal}\sqrt{\rho_{input}}})^2$, we calculated the fidelity of the reconstructed density matrix by comparing it with the ideal density matrix, which was of 91.0%±1.8%. Where $\rho_{ideal}$ was the density matrix corresponding to the ideal OAM entangled state of $\Psi=(|L>|L>+|R>|R>)/2^{1/2}$, $\rho_{input}$ was the reconstructed density matrix. Next we sent the signal 1 photon to MOT 2 for subsequent storage via FORTPT. After a programmed storage time of 150 ns, the signal 1 was retrieved by swiching on the coupling light again. Then we applied the pump 2 to map the spin excited state of the ensemble in MOT 1 to signal 2 photon. We measured the time-correlated function between the signal 2 photon and the retrieved signal 1 photon, reconstructed the density matrix according to the measured coincidence rates. The real/imaginary part of the reconstructed density matrix of the OAM entanglement was shown by Fig. 3(c)/(d). The fidelity of the density matrix calculated was 84.6%±2.6% by comparing it with the ideal density matrix $\rho_{ideal}$, and was 90.3%±0.8% compared with the reconstructed density matrix $\rho_{input}$ before storage by using the formula of $F_2 = Tr(\sqrt{\sqrt{\rho_{output}}\rho_{input}\sqrt{\rho_{output}}})^2$, where $\rho_{output}$ was the reconstructed density matrix after storage.

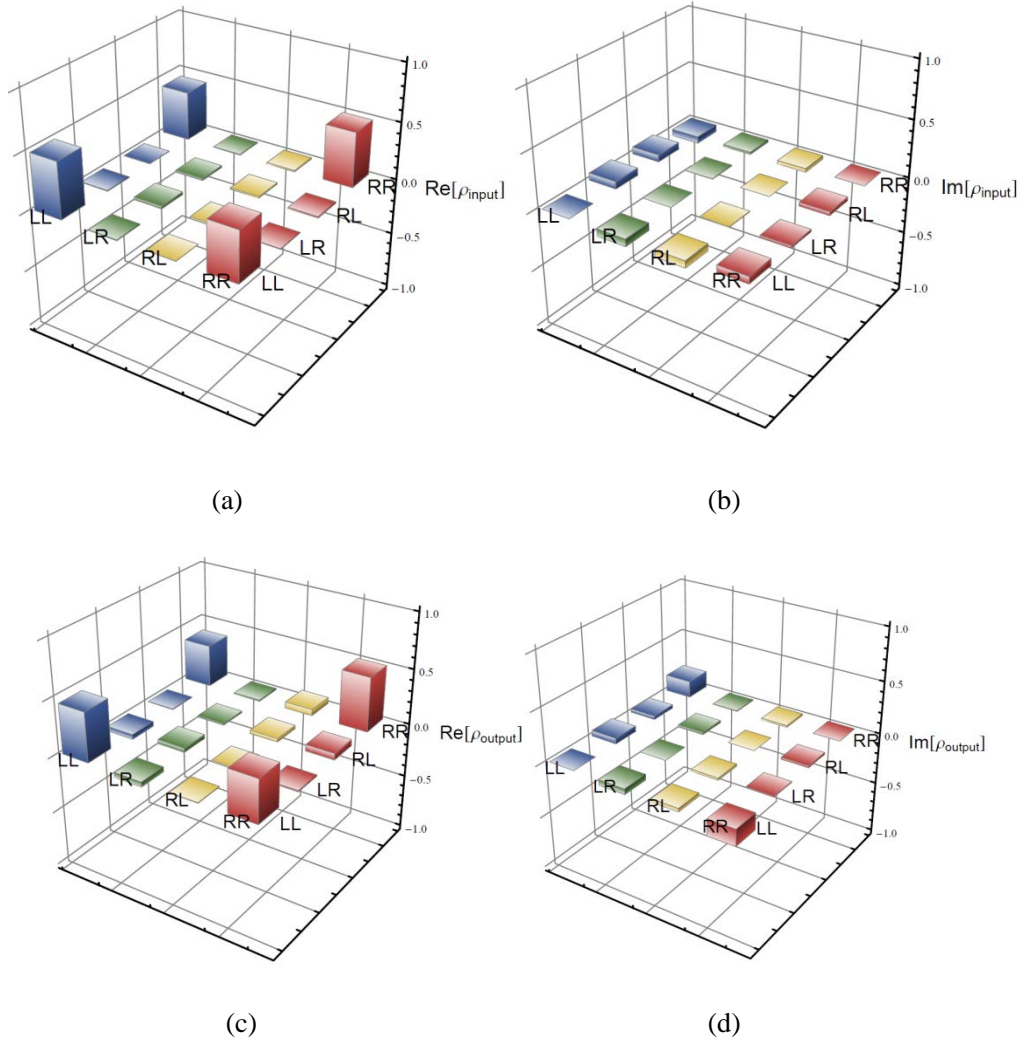

(a)　　　　　　　　　　　　(b)

(c)　　　　　　　　　　　　(d)

**Figure.3|.The reconstructed density matrices before and after storage.**(a) and (b) are the real and imaginary parts of the reconstructed density matrix of the state before storage respectively. (c) and (d) correspond to the real and imaginary parts of the reconstructed density matrix of the state after storage respectively. The background noise has been subtracted. The background noise was estimated by repeating the experiment without input signal 1 photon to MOT 2. The measurement time for each data was 500 s in (a) and (b) and 1000 s in (c) and (d).

We further characterized the degree of entanglement after storage through checking the Bell's inequality, which was the symmetrized version called CHSH inequality [48-50]. For our experiment, the CHSH parameter $S$ was referenced by Ref. [51] as:

$$S = E(\theta_A, \theta_B) - E(\theta_A, \theta_B') + E(\theta_A', \theta_B) + E(\theta_A', \theta_B') . \quad (1)$$

Where $\theta_A$, $\theta_B$ were the angles of the phase distributions on the surfaces of SLMs which were defined in Fig. 6 (b) of Methods. $E(\theta_A, \theta_B)$ could be calculated from the coincidence rates at particular orientations,

$$E(\theta_A,\theta_B) = \frac{C(\theta_A,\theta_B)+C(\theta_A+\frac{\pi}{2},\theta_B+\frac{\pi}{2})-C(\theta_A+\frac{\pi}{2},\theta_B)-C(\theta_A,\theta_B+\frac{\pi}{2})}{C(\theta_A,\theta_B)+C(\theta_A+\frac{\pi}{2},\theta_B+\frac{\pi}{2})+C(\theta_A+\frac{\pi}{2},\theta_B)+C(\theta_A,\theta_B+\frac{\pi}{2})}$$

In our experiment, we selected $\theta_A=0$, $\theta_B=\pi/8$, $\theta_A'=\pi/4$, $\theta_B'=3\pi/8$. The calculated $S$ was of $S=2.48\pm0.04$ before storage and $S=2.41\pm0.06$ after storage. The inequality is violated when the values of $S$ is greater than 2 and the violation of inequality means that there exists the entanglement between photons. These results experimentally obtained clearly showed that the CHSH inequality was violated even after a programmed storage time, demonstrating the preservation of the OAM entanglement during the storage.

Moreover, we checked the two-photon interference. As we knew that if the visibility of the two-photon interference was >70.7%, then CHSH inequality would be violated, proving the entanglement existed between two photons. In this experiment, we fixed the phase angleof SLM 1 to be $\theta_A=0°$ or 45° respectively, measured the coincidence rate at different angles of $\theta_B$ of the SLM 2. The storage time was set to be of 150 ns. The experimental results were shown by Fig. 4, visibilitywas 85.2%±3.0% at $\theta_A=0°$ and 86.8%±3.3% at $\theta_A=45°$, both were larger than 70.7%, clearly proved the preservation of the OAM entanglement in storage again.

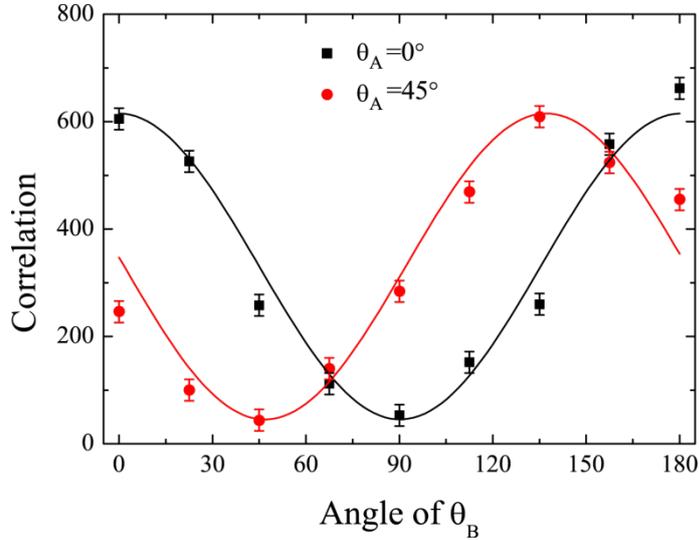

**Figure.4|.The measured coincidence rate at $\theta_A=0°$ and 45° with different $\theta_B$.** The red (blue) curve represents the correlated coincidence rate with the orientations of $\theta_A=0$ (45°). The background noise has been subtracted. Error bar is ±1 standard deviation. The background noise was estimated by repeating the experiment without input signal 1 photon to MOT 2. The measurement time for each data was 1000 s.

**Discussions**

We want to mention that we can prove the storage of a high-dimensional OAM entangled states (*d*≥3) by reconstructing the density matrix like done in two-dimensional case in principle, but in practice, there are some big challenges in the experimental realization: for example, how to achieve higher signal-to-noise ratio and how to stabilize the system over long period. This is because we have to take more data in case of storing a high-dimensional entanglement. For example, reconstructing a process density matrix of quantum memory for a three-dimensional state (*d*=3) requires much longer experimental times owing to the necessary measurement of 81 data points [52].

This memory could store a high-dimensional OAM entangled states (*d*≥3) in principle, the number of dimension of per photon can be simply estimated by the formular of $w(z) = \sqrt{l+1}w_0(z)$, where, w(z) is the beam waist of a light carrying OAM of *l* at the center of the atomic vapor in MOT 2, and $w_0(z)$ is the beam waist of a Gaussian light. In our experiment, $w_0(z) \sim$ 100 µm, and the radius of the atomic vapor in MOT 2 is ~1 mm, therefore *l*~100, i.e., the maximal OAM dimension of per photon could be stored in our system was limited to be 200. Of course, this number is also limited by other factors, such as the Fresnel number and the optical depth of the atomic ensemble [53], the angle of signal field and the control field in MOT 2, etc., which need further investigation.

The fidelity of the measuremented OAM entangled state prepared in MOT 1 was below 100%, one main reason was from the imperfect measurement: in the experiment, the distance between the SLM 1 and SLM 2 was 3 m. The mismatch between the positions of these two SLMs where the signal photons should be entangled and the actual positions reduced the fidelity. One could soften this influence by enlarging the beam waist of the signal 1 and signal 2 photons. Another reason was from the dephasing between the two ground states induced by Earth's magnetic field.

**Conclusion**

In conclusion, we had reported the first experimental realization of storing OAM entanglement via FORTPT in an atomic ensemble. The fidelities of the reconstructed density matrices in a postselected two-dimensional subspace before and after storage were 91% and 84% respectively compared with the ideal density matrix. In addition, the violations of CHSH inequality were experimentally demonstrated before and after storage. Besides, we also realized the first experimental storage of a true-single-photon state via FORTPT in an atomic ensemble, and also provided an experimental evidence that an image memory at single-photon level can be realized via FORTPT too. This work clearly demonstrates the workability of building up a high-dimensional quantum communication system in the future.

## Methods

**Storing a weak light via FORTPT in a cold atomic ensemble system.**

In order to further illustrate the memory via FORTPT, we showed the spectrum of transmission of the signal 1 through the atomic ensemble in MOT 2 by using a weak coherent light. There was a pumping hole in the spectrum of transmission of signal 1 at the detuning of 70 MHz. The power of coupling laser was 12 mW. The experimental obtained curve was shown in Fig. 5(a). Then we stored the weak coherent light via FORTPT in the atomic ensemble in MOT 2, the experimental data were shown by Fig. 5(b) and 5(c).

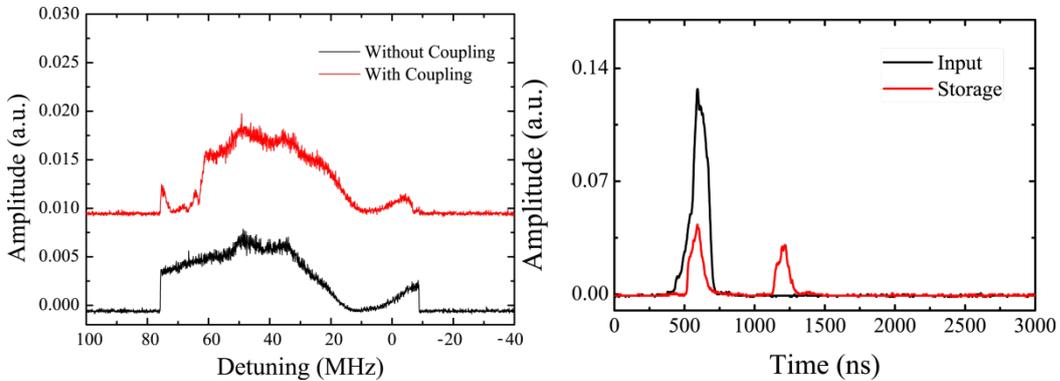

(a)                                         (b)

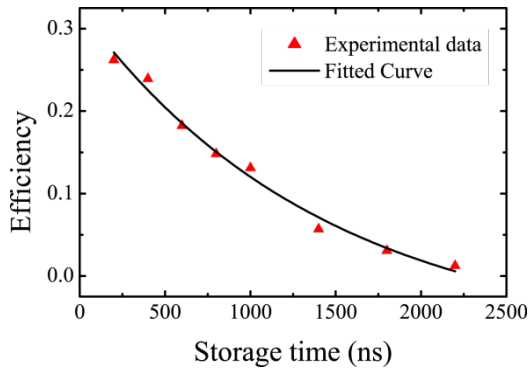

(c)

**Figure.5|.The storage of a weak coherent light via FORTPT.**(a) The transmission spectrum of a weak coherent light through the atomic ensemble in MOT 2. Black line is the transmission spectrum without coupling laser. Red line represents the transmission spectrum with coupling laser.(b) The storage of a weak coherent light via FORTPT with the time of 400 ns. (c) The memory efficiency against the storage time.The solid line is an exponential fit by $g_0+Ae^{-(\tau-\tau_0)/T}$ ($g_0$=-0.08;A=0.38;$\tau_0$=67; T=1434).

**Experimental preparation of the nonclassical correlation in time domain betwee two atmoic ensembles**

In the experiment shown Fig. 1, pump 1 was applied firstly for generating an anti-Stokes photon at 795 nm called signal 1 by a non-colinear SRS process in MOT 1. The optical depth (OD) of MOT 1 measured was about 8. The generated signal 1 photon was delivered into the second MOT 2 for subsequent storage. The OD of MOT 2 was 20. A coupling pulse laser with orthogonal polarization to signal 1 was used to store the signal 1 via FORTPT. There was an angle of 3º between the signal 1 and coupling laser for reducing the scattering noise from the coupling laser. After a programed storage time, the signal 1 photon was retrieved and reflected by a spatial light modulator (SLM 2, HOLOEYE, PLUTO. The active area is of 15.36 mm × 8.64 mm, the pixel pitch size is 8μm, and with a resolution of 1920×1080. The reflectivityis of 60%.) and then coupled into a single mode fibre (with an overall efficiency of 30%, including the loss of lens, mirrors, mirror of the MOT 2 and the coupling efficiency of the single mode fibre). After the signal 1 photon was retrieved from MOT 2, the pump 2 laser, which counter-propagated through MOT 1 with pump 1, was switched on for converting the collective spin excited state of the atomic ensemble in MOT 1 to a Stokes photon called signal 2, whose wavelength was 780 nm. Pump 1 and pump 2 had the opposite linear polarizations. The singal 1 and signal 2 counter-propagated and the angle between the signal 1 and pump 2 beams was set to be 3 º in order to reduce the scattering noise from the pump lasers. The signal 2 photons were input onto the surface of another spatial light modulator (SLM 1), and the reflected photons were coupled into another single mode fibre with a coupling efficiency of 80%. Both photons were detected by two single photon detectors 1 and 2 respectively (PerkinElmer SPCM-AQR-15-FC, with an efficiency of 60%), whose outputs were input to a time-to-digital converter (Fast Comtec P7888) with 1 ns bin width for coincidence measurement. The signal 1 acted as trigger photon, and the signal 2 acted as stop signal. The pump 2 was filtered by an interference filter (with a transmission rate of 95%) for reducing the fluorescence noise from diode laser. Several Fabry-Perot etalons with a bandwidth of 500 MHz each were inserted into the optical routes of signal 1 and signal 2 for reducing the noises further, the overall transmission rates were 50% and 60% respectively. The experiment was performed with a repeat rate of 100 Hz. In each cycle, the time for atomic trapping was 8 ms, the experimental window was 1.5 ms, and another 0.5 ms was used to initialize the atomic state.

The signal 1 and signal 2 photons were correlated in time domain, which could be proved by checking whether the Cauchy-Schwarz inequality is violated. Usually classical lights satisfy the equation of $R = \frac{[g_{s1,s2}(\tau)]^2}{g_{s1,s1}(0)g_{s2,s2}(0)} \leq 1$. If R>1, the light is non-classical. Where $g_{s1,s2}(\tau)$, $g_{s1,s1}(0)$, and $g_{s2,s2}(0)$ are the normalised second-order cross-correlation and auto-correlation of the photons respectively. The normalized $g_{s1,s2}(\tau)$ can be obtained by normalizing the true two-photon coincidence counts to the accidental two-photon coincidence counts $g_{s1,s2}(\infty)$. (Since the time-correlated function in pulse-mode was comb-like peak structure, $g_{s1,s2}(\tau)$ was calculated by the ratio between the coincidence rate in the first peak and that in the second one.) With $\tau=t_{s1}-t_{s2}$, the relative time delay between paired photons, the maximum $g_{s1,s2}(\tau)$ obtained was $g_{s1,s2}(\tau)$=12 at $\tau$=380 ns with the storage time of 150 ns. Thus, R=38, was much larger than 1 using the fact that $g_{s1,s1}(0)=g_{s2,s2}(0)\approx2$ (photons from signals 1 and 2 exhibited photon statistics typical of thermal light), the Cauchy-Schwarz inequality was strongly violated, clearly demonstrating the preservation of non-classical correlation during the storage. The bandwidth of the photon was ~20 MHz.

We also characterized the single photon property of the signal 1 photon by checking an anti-correlation parameter $\alpha$ ($\alpha = \frac{P_1 P_{123}}{P_{12} P_{13}}$, where $P_1$ was the signal 2 photon counts, $P_{12}$ and $P_{13}$ were the twofold coincidence counts between the signal 2 photon and the two separated signal 1 photon by a beam splitter, $P_{123}$ was the threefold coincidence counts), which were of 0.074±0.012 before storage and 0.29±0.02 after 150 ns-long storage. For an ideally prepared single-photon state, *α* tends to zero, for a classical field, *α*≥1, based on the Cauchy-Schwarz inequality. A pure single photon has α=0 and a two-photon state has α=0.5. Therefore α<1.0 violates the classical limit and α<0.5 suggests the near-single-photon character. Both went to zero confirmed clearly the preservation of the single-photon nature in storage. The experimental results also clearly demonstrated the successful storage of a true-single-photon via FORTPT in an atomic ensemble.

The efficiency of storage was 26.7%. Dephasing between the two ground states induced by Earth's magnetic field had an effect during storage, which shortened the storage time and reduced the storage efficiency. Besides, the efficiency was also restricted by coupling laser power [44, 45]. A coupling laser with higher power could improve the effieicncy further.

**Tomography of entanglement and the demonstrtation of violation of CHSH inequality.**

AS said in the main text, in order to simplify the experiment, shorten the measurement time, we only experimentally demonstrated the storage of the OAM entanglement postselected in a two-dimensional subspace. We considered the quanta of OAM $l=±1\hbar$. The OAM photons was entangled in an expression of $\Psi=(|L>|L>+|R>|R>)/2^{1/2}$, where $|L>$, and $|R>$ were states corresponding to a well-defined OAM of $1\hbar$ and $-1\hbar$ respectively. To reconstruct the density matrix of the entangled OAM state, both photons were input to two SLMs respectively, where four states of $|\psi_{1\sim4}>(|L>, |R>, (|L>+|R>)/2^{1/2}, (|L>-i|R>)/2^{1/2})$ shown by Fig. 6 (a) were programed onto the SLM 1 and SLM 2 by two computers. The reflected photons from SLMs were collected into two single-mode fibers respectively. We measured each correlated coincidence, obtained a set of 16 data and used them to reconstruct the density matrix. The error bar of our experimental measurement was estimated from Poissonian statistics and using Monte Carlo simulations.

In order to check the CHSH inequality and observe two-photon interference, we defined the different sector states bysetting different angle phase distributions onto the SLMs shown by Fig. 6(b), and measured the corresponding coincidence rate with these states. Using these coincidence rates, we calculated the CHSH inequality in the cases of before and after storage. In experiment, the conjugate properties owing to the 4-f imaging system used had be considered when the signal 2 was input onto SLM 2. The details referred to Ref. [53]. Fig. 6(c) is the details of our experimental setup. There were one and five mirrors (four mirrors plus a PBS) mounted in optical routes of signal 2 and signal 1 photons respectively. Thus, our system was symmetrical because of the odd-numbered mirror in each optical route. If the mirror 2 (M2) was removed, the entangled measurement will be of $\Psi=(|L\rangle|R\rangle+|R\rangle|L\rangle)/2^{1/2}$. Another point was that the 4-f imaging system was constructed by lens 1 and lens 2, which induced the inverted imaging in the signal 1 photon.

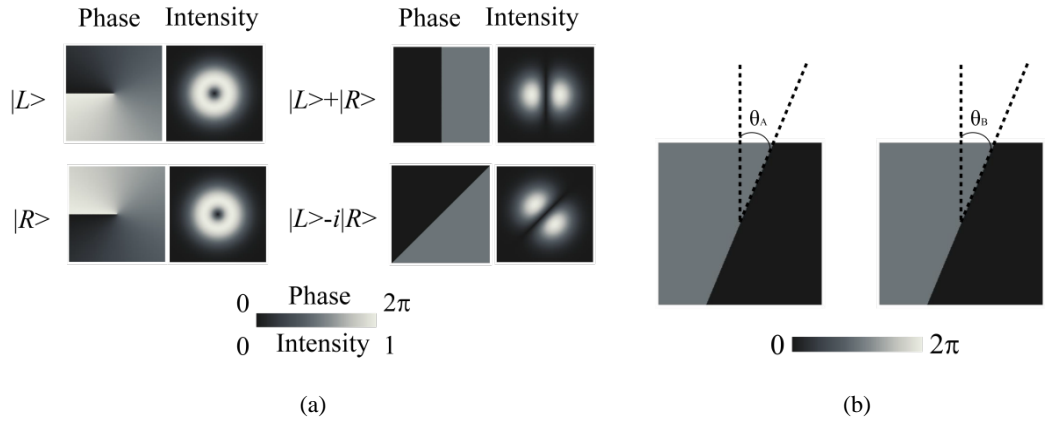

(a)                  (b)

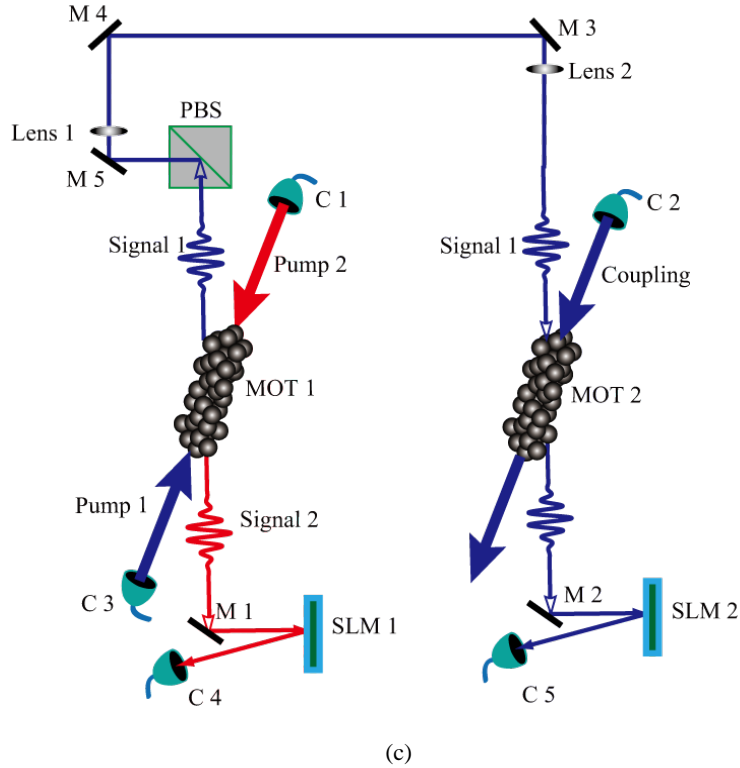

(c)

**Figure.6|.**(a) The four OAM states for reconstructing density matrix.The phase (left panel of each pair) and the intensity (right panel) distributions of four OAM states. All images are explanation of image measurement, in practical the images we used are the transformed due to same optical components, such as mirrors and lens. (b) The different OAM superpositionstates defined different angles $\theta_A$, $\theta_B$. (c) The details of experimental setup. Lens 1 and Lens 2 consisted of a 4-f imaging system. C: fibre coupler; M: mirror. MOT: magneto-optical trap; SLM: spatial light modulator; PBS: polarisation beam splitter.

**Author contributions**

BSS and DSD conceived the experiment. The experimental work and data analysis were carried out by DSD and BSS, with assistance from ZYZ, WZ and SS. BSS and DSD wrote this paper with assistance from ZYZ, GYX, YKJ and XSW. BSS and GCG supervised the project.

**Acknowledgements**

We thank Dr.Yun-Feng Huang and Dr. Jin-Shi Xu for helpful discussions.This work was supported by the National Fundamental Research Program of China (Grant No. 2 011CBA00200)**,** the National Natural Science Foundation of China (Grant Nos. 11174271, 61275115), the Youth Innovation Fund from USTC (Grant No.ZC 9850320804), and the Innovation Fund from CAS.